\documentclass[apjl]{emulateapj}
\usepackage{apjfonts}

\begin{document}

\title{The dramatic size evolution of elliptical galaxies and the quasar feedback}
\author{L. Fan\altaffilmark{1,2}, A. Lapi\altaffilmark{3,1}, G. De Zotti\altaffilmark{4,1}, and L.
Danese\altaffilmark{1}} \altaffiltext{1}{Astrophysics Sector,
SISSA/ISAS, Via Beirut 2-4, 34014 Trieste, Italy}
\altaffiltext{2}{Center for Astrophysics, University of Science
and Technology of China, Hefei, 230026, China}
\altaffiltext{3}{Dip. Fisica, Univ. ``Tor Vergata'', Via
Ricerca Scientifica 1, 00133 Roma, Italy}
\altaffiltext{4}{INAF-Osservatorio Astronomico di Padova,
Vicolo dell'Osservatorio 5, 35122 Padova, Italy}

\begin{abstract}
Observations have evidenced that passively evolving massive
galaxies at high redshift are much more compact than local
galaxies with the same stellar mass. We argue that the observed
strong evolution in size is directly related to the quasar
feedback, which removes huge amounts of cold gas from the
central regions in a Salpeter time, inducing an expansion of
the stellar distribution. The new equilibrium configuration,
with a size increased by a factor $\gtrsim 3$, is attained
after $\sim$40 dynamical times, corresponding to $\sim 2$ Gyr.
This means that massive galaxies observed at $z\geq 1$ will
settle on the Fundamental Plane by $z\sim 0.8$--1. In less
massive galaxies ($M_{\star}\lesssim 2 \times
10^{10}\,M_{\odot}$), the nuclear feedback is subdominant, and
the mass loss is mainly due to stellar winds. In this case, the
mass loss timescale is longer than the dynamical time and
results in adiabatic expansion that may increase the effective
radius by a factor of up to $\sim 2$ in 10 Gyr, although a
growth by a factor of $\simeq 1.6$ occurs within the first 0.5
Gyr. Since observations are focused on relatively old galaxies,
with ages $\gtrsim 1\,$Gyr, the evolution for smaller galaxies
is more difficult to perceive. Significant evolution of
velocity dispersion is predicted for both small and large
galaxies.
\end{abstract}


\section{Introduction}
Several recent observational studies have found that massive,
passively evolving, galaxies at $z>1$ are much more compact
than local galaxies of analogous stellar mass (Ferguson et al.
2004; Trujillo et al. 2004, 2007; Zirm et al. 2007; Cimatti et
al. 2008; Damjanov et al. 2008). Since similarly superdense
massive galaxies are extremely rare or absent at $z\simeq 0$
(Shen et al. 2003) a strong size evolution, by a factor $\sim
3$ or more, is indicated. No convincing mechanism able to
account for such size evolution has been proposed so far. In
the following we show that an expansion consistent with the
observed one is naturally expected as a consequence of feedback
from active nuclei (Silk \& Rees 1998; Granato et al. 2001,
2004), which is now widely recognized as a crucial ingredient
of semi-analytic models (see, e.g., Di Matteo et al. 2008).

In the local Universe spheroidal galaxies occupy a quite narrow
region, the Fundamental Plane, in the 3-dimensional space
identified by the effective or half-light radius $r_e$, by the
central velocity dispersion $\sigma_0$, and by the mean surface
brightness within $r_e$ (Djorgovski \& Davis 1987). The tight
color-magnitude relation, the color-velocity dispersion
relation, and spectral line indices imply that the bulk of
stars of elliptical galaxies formed at $z\geq 1.5$. The
enhancement of $\alpha$-elements abundances with respect to
iron in massive elliptical galaxies entails that most of their
stars formed within the first Gyr of their life (see Renzini
2006 for a comprehensive discussion).

An additional key result is the generic presence of a Super
Massive Black Hole (SMBHs) in the center of local elliptical
galaxies. Its mass is directly proportional to the mass of the
old stellar population, $M_{\rm BH}\sim 2\times
10^{-3}\,M_{\star}$ (Magorrian et al 1998; see Ferrarese \&
Ford 2005 for a review), implying that quasars and spheroidal
galaxies form and evolve in strict relation and with mutual
feedback. Specifically, it has been suggested that the central
SMBH  grows until until its feedback unbinds the residual gas
in the host galaxy and sweeps it out through a high velocity
wind, thus halting both the star formation and its own fueling
and establishing a relationship between the SMBH mass and the
stellar velocity dispersion (Silk \& Rees 1998). Strong AGN
feedback also appears to be the only viable mechanism to
explain the exponential cut-off at the bright end of the galaxy
luminosity function (Croton et al. 2006) and the observed
bimodality in the color-magnitude diagram of galaxies at
$z\gtrsim 1.5$ (Menci et al. 2006). Direct observational
indications of massive outflows close to high redshift quasars,
consistent with this scenario, have been reported (e.g. Simcoe
et al. 2006; Prochaska \& Hennawy 2008).

As shown below, the amount of gas rapidly stripped from the
central regions of the galactic halo can be large enough to
drive a large increase of the galaxy size. The puffing up of a
system by rapid mass loss is a well known phenomenon,
extensively studied both analytically and through numerical
simulations, with reference to galaxies (Biermann \& Shapiro
1979), and, especially, to globular clusters (Hills 1980;
Goodwin \& Bastian 2006). Slower, adiabatic expansion is caused
by mass loss due to stellar winds or supernova explosions
(Hills 1980; Richstone \& Potter 1982).

In this {\it Letter} we first summarize the effect of mass loss on the galaxy
size evolution (\S\,2), then we present quantitative estimates on the evolution
of the effective radius and of the central stellar velocity dispersion as a
function of galactic mass, using the Granato et al. (2004) model as a reference
(\S\,3), and finally, in \S\,4, we summarize and discuss our results.

\section{Size evolution due to mass loss from virialized systems}\label{sect:loss}

The effect of the mass loss on the structure and dynamics of a
virialized stellar system depends on the amount of ejected mass
and on the timescale of ejection. Two regimes can be
identified, corresponding to an ejection timescale, $\tau_{\rm
ej}$, shorter or longer than the dynamical timescale $\tau_{\rm
dyn}$.

\subsection{Rapid mass loss}\label{sect:rapid}

A rapid mass loss ($\tau_{\rm ej}< \tau_{\rm dyn}$) results in
a shallower potential well at essentially constant velocity
dispersion. If $M$ and $M'$ are the initial and final masses,
the final energy, $E'$, of a spherical system is related to the
initial energy $E$ by $E'=E(M'/M)^2(2-M/M')$ (Biermann \&
Shapiro 1979), so that if $M/M'>2$ the system has positive
energy and is unbound. If $M/M'<2$, the system will eventually
relax to a new equilibrium configuration. If the latter is
homologous to the initial one, the ratio of initial ($R$) to
final ($R'$) radii is
\begin{equation}\label{eq:fast}
R/R'= 2-M/M'\ .
\end{equation}
Numerical simulations essentially confirm this simple estimate
and show that, if the system is not disrupted, the new
equilibrium configuration is reached after about 30--40 initial
dynamical times (see, e.g., Goodwin \& Bastian 2006).

It is important to note that the above argument assumes that
the system is self-gravitating and isolated. Clearly the dark
matter (DM) halo, which extends far beyond the stellar
distribution, exerts a stabilizing action and prevents its
disruption.

\subsection{Long lasting, adiabatic mass loss}

If the mass loss occurs on a timescale $\tau_{\rm ej}>
\tau_{\rm dyn}$, the system expands through the adiabatic
invariants of the orbits of stars, and the expansion proceeds
at a rate proportional to the mass loss rate (Hills, 1980;
Richstone \& Potter 1982). We have:
\begin{equation}\label{eq:slow}
R'/R= M/M'\ .
\end{equation}
In this case no  global disruption of the system is possible
and the velocity is inversely proportional to the size.

\section{Mass loss and size evolution of high redshift galaxies}\label{sect:size}

Observations of (sub)-millimeter bright QSOs, thought to be close to the onset
of the massive outflows determining the transition from the active star-forming
phase to the unobscured QSO phase, indicate very large gas masses in the
central regions of their host galaxies (Lutz et al. 2008, and references
therein), comparable to the stellar masses (Coppin et al. 2008). Similar star
to gas ratios are found in the central regions of sub-mm bright galaxies
(Tacconi et al. 2008).

About 5\% of the QSO bolometric luminosity during the maximum quasar activity
is enough to remove the gas in about one Salpeter time ($\simeq 4\times
10^7\,$yr), through outflows of $\dot{M}\gtrsim 1,000\,
M_{\odot}\,\hbox{yr}^{-1}$ (e.g. Granato et al. 2004). The high precision
spectroscopy of the QSO SDSSJ$1204+0221$ by Prochaska \& Hennawi (2008) is
indeed consistent with a high velocity ($v\simeq
1,000\,\hbox{km}\,\hbox{s}^{-1}$), massive outflow of $\dot{M}\sim 3,000\,
M_{\odot}\, \hbox{yr}^{-1}$.

There are thus both observational indications and theoretical
arguments for the occurrence of the conditions leading to the
expansion described in \S\,\ref{sect:rapid}. For sake of
definiteness, we will present quantitative estimates exploiting
the model by Granato et al. (2004), which reproduces the
crucial observational features of the coevolution of quasars
and elliptical galaxies (see Lapi et al. 2006 for a
comprehensive discussion). In this model, both the star
formation and the growth of the SMBH proceed faster in more
massive halos, in keeping with the observed ``downsizing''. In
the most massive galaxies, the quasar feedback quenches the
star formation after $\sim 0.5\,$Gyr, thus accounting for the
observed $\alpha$-enhancement. The star formation occurs in the
inner regions of the galactic halo, encompassing a fraction
$\leq 20$--$30\%$ of the baryons associated to the halo. The
star formation may be triggered and enhanced by the rapid
merging of a few sub-halos. The dynamical friction is able to
dissipate angular momentum of the collapsing baryons,
transferring it to DM particles. The numerical simulations by
El-Zant et al. (2001) showed that dynamical friction between
baryon and DM clouds within the scale radius of the Navarro et
al. (1997) density distribution ($r\leq r_s$) is strong enough
to overcome the expected adiabatic contraction of DM caused by
the deepening of the potential well under the influence of the
shrinking baryons. Eventually the central regions are
gravitationally dominated by baryons (stars and cold gas), that
may share the velocity field of the DM clouds (cfr. Fig. 3 of
El-Zant et al. 2001). Under more general assumptions, the
baryon collapse is expected to result in stellar velocity
dispersions somewhat higher than those of the DM. We set
$\sigma_{\star}\approx f_{\sigma}\sigma_{\rm DM}$.

We can assume that the stars and the cold gas, of mass
$M_{\star}$ and $M_{\rm cold}$ respectively, are in virial
equilibrium just before the mass loss. Neglecting the DM
contribution to the mass in the central region, its
gravitational radius can be expressed as $r_g\approx G
\left(M_{\star} + M_{\rm cold}\right)/\sigma_{\star}^2$ and the
dynamical time scale of the collapsed baryon component is
$\tau_{\rm dyn}=(\pi/2)(r_g/\sigma_{\star})$.

If we adopt a S\'ersic's (1968) law  for the projected stellar
density profile, the effective radius, that we identify with
the projected half stellar mass radius, is related to the
gravitational one by $r_e=S_s(n) r_g$, and the observed
luminosity weighted {\it central} velocity dispersion is
related to the global density-weighted velocity dispersion by
$\sigma_0^2=S_K(n) <\sigma^2>$, where $n$ is the S\'ersic index
($n=4$ for a de Vaucouleurs (1959) $r^{1/4}$ law). Both
$S_s(n)$ and $S_K(n)$ are rather weak function of $n$ tabulated
by Prugniel \& Simien (1997); in particular, $S_s(4)\simeq
0.34$ and $S_K(4)\simeq 0.52$.

For a Navarro et al. (1997) profile,  the halo circular
velocity, $V_H$, and the virial radius $r_{\rm vir}$ are well
approximated by $ V_H\simeq 230 (M_H/10^{12} M_{\odot})^{1/3
}[(1+z)/4]^{1/2}\,\hbox{km}\,\hbox{s}^{-1}$ and $r_{\rm
vir}=80(M_H/10^{12} M_{\odot})^{1/3 }[(1+z)/4]^{-1}
\,\hbox{kpc}$, respectively. If the density profile has a scale
length $r_s$, the DM velocity dispersion is $\sigma_{\rm DM}=
f(c)^{1/2}V_H$, $f(c)\approx 1$ being a weak function of the
concentration $c=r_{\rm vir}/r_s$. Rearranging, we obtain the
dynamical time and the effective radius as a function of virial
radius and mass of the host halo, and of the mass in stars and
in cold gas at the onset of the mass loss:
\begin{equation}\label{eq:re}
r_e\approx \frac {S_s(4)}{f_{\sigma}^2} \frac{\left(M_{\star}+M_{\rm cold}\right)}{M_H}  r_{\rm vir} \ .
\end{equation}
The model gives the mass in stars and in gas, and the mass loss for any choice
of the galaxy halo mass, $M_H$, and of the virialization redshift. We can then
compute the effective radius with eqs.~(\ref{eq:re})  and its evolution with
eqs.~(\ref{eq:fast}) and (\ref{eq:slow}). The application of the virial
theorem, using the coefficients $S_s(4)$ and $S_K(4)$, allows us to derive the
central line-of-sight velocity dispersion in different evolutionary phases. The
results are shown in Fig.~\ref{fig:sigma}.

Simple analytical approximations for the mass in stars and cold
gas as a function of halo mass and redshift were derived Mao et
al. (2007, Appendix A) in the framework of the Granato et al.
(2004) model. The key timescales of the model are the
condensation time of the cold gas out of the hot phase at
virial temperature, $t_{\rm cond}\approx 9\times 10^8
[(1+z)/4]^{-1.5} (M_H/10^{12} M_{\odot})^{0.2}\,$yr, and the
duration of the star formation phase before the onset of the
quasar super-wind, $\Delta t_{\rm burst}\approx 6\times 10^8
[(1+z)/4]^{-1.5} F(M_H/10^{12} M_{\odot})\,$yr, where $F(x)=1$
for $x\geq 1$ and $F(x)=x^{-1}$ for $x<1$. The mass in stars
formed during the burst is
\begin{equation}
M_{\star}/M_H\sim s\frac {f_b}{s\gamma-1}[1-\exp(-\Delta t_{\rm
burst}/t_{\rm cond})] ,
\end{equation}
where $f_b\simeq 0.18$ is the cosmic baryon to DM density ratio
and the function $\gamma \simeq 1-R+0.6
(M_H/10^{12}M_{\odot})^{-2/3 }[(1+z)/4]^{-1}$ incorporates the
supernova feedback effect. The ratio between the condensation
time and the characteristic time of star formation $s=t_{\rm
cond}/t_{\star}\approx 5$ accounts for the clumping of the gas
distribution; $R\simeq 0.3$ denotes the fraction of mass
restituted by dying stars to the ISM. During the last e-folding
time the quasar activity ejects the residual cold gas, which
amounts to
\begin{equation}
\frac {M_{\rm cold}}{M_{\star}} \simeq \frac {M_{\rm
ej}}{M_{\star}}\approx [ s  \left (\exp(\Delta t_{\rm
burst}/t_{\rm cond})-1\right )]^{-1} .
\end{equation}
For massive galaxies hosted in halos of $M_H\gtrsim 10^{12}\,
M_{\odot}$, $M_{\rm cold}/M_{\star}\sim 2/3$. The ejection of a
mass $\simeq M_{\rm cold}$ can then produce a factor $\simeq 3$
increase in galaxy size (cfr. eq.~(\ref{eq:fast})). On the
other hand, for smaller galaxies ($M_H < 10^{12} \,M_{\odot}$
and present day stellar mass $M_{\star} < 2\times 10^{10}\,
M_{\odot}$) the nuclear activity is much weaker and occurs
later, when the ratio $M_{\rm cold}/M_{\star}$, and hence
$M_{\rm ej}/M_{\star}$, is small. This is in keeping with the
general picture that the early evolution of small galaxies is
ruled by SN feedback, while the evolution of the large ones is
controlled by quasar feedback (see, e.g., Shankar et al. 2006).

\begin{figure}
\epsscale{1}\plotone{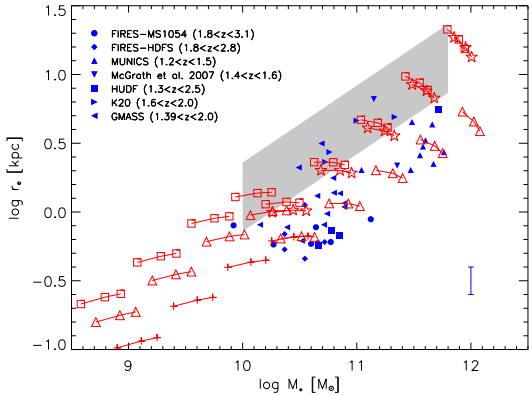}\caption{Effective radius versus
stellar mass. For $M_\star\geq 2\times 10^{10}\,M_{\odot}$
triangles, asterisks, and squares show, respectively, the
values of $r_e$: at the end of the bright quasar phase;
when the system has reached a new virial equilibrium; at the present time. For $M_\star\leq
2\times 10^{10}\,M_{\odot}$ crosses, triangles, and squares
show, respectively, the values of $r_e$: just at the end of
baryon collapse; after further 0.5 Gyr;  at
the present time. The three connected symbols
refer to different halo virialization redshifts ($z_{\rm
vir}=2$, 4, and 6); smaller values of $r_e$ correspond to
higher values of $z_{\rm vir}$. The grey area marks the $\pm 1
\sigma$ local size--$M_\star$ relation (Shen et al. 2003). Data
points as in Fig.~15 of Cimatti et al. (2008). A typical error bar is shown in the
lower right-hand corner.} \label{fig:re}
\bigskip
\end{figure}

\begin{figure}
\epsscale{1}\plotone{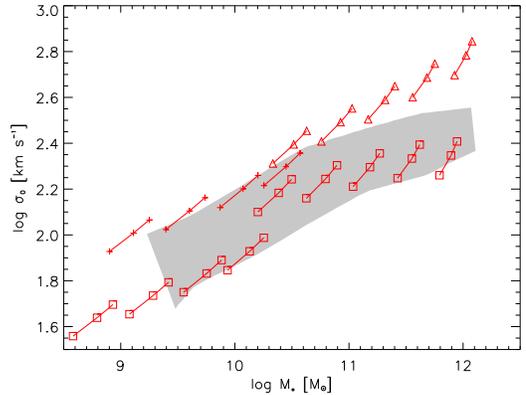}\caption{Evolution of the
central line-of-sight velocity dispersion versus stellar mass.
Symbols as in Fig.~\protect\ref{fig:re};
the same values of $z_{\rm vir}$ and $f_{\sigma}$ as in
Fig.~\protect\ref{fig:re} are adopted. The grey area reproduces
the distribution of local data (Gallazzi et al. 2006).}
\label{fig:sigma}
\end{figure}

\section{Discussion and Conclusions}

The variation of the effective radius  implies that elliptical
galaxies spend the initial part of their life outside the
'local' fundamental plane and only subsequently settle on it.

For massive galaxies ($M_\star\geq 2\times 10^{10}\,M_{\odot}$), for which the
expansion is mostly due to quasar-driven super-winds, Fig.~\ref{fig:re}
displays three steps in the evolution: {\it i)} the end of the quasar phase at
cosmic time $t_{\rm QSO}\simeq t_{\rm vir}+\Delta t_{\rm burst}$, which
corresponds to the end of the rapid mass loss; {\it ii)} the intermediate time
$t_{\rm int}$ when the stellar component reaches a new virial equilibrium with
a larger size; {\it iii)} the present time $t_0$, at which a further, minor
increase of the size is achieved due to stellar winds. The initial effective
radius has been computed from eq.~(\ref{eq:re}) with $f_{\sigma}\approx 1.3$.

Equation~(\ref{eq:fast}) shows that the size expansion can be
extremely large as the fraction of expelled gas approaches 50\%
of the initial mass. However, at variance with the case of star
clusters, confined by their own gravitation field, the presence
of the DM halo, dynamically dominant at large radii, prevents
the disruption and restricts the range of possible final
structures of large elliptical galaxies.

The time lapse required to reach the new virial equilibrium can be inferred
from numerical simulations of stellar clusters, which show that the system
reaches the new equilibrium after about 40 dynamical times (computed for the
initial configuration), independently of the initial mass (Geyer \& Burkert
2001; Goodwin \& Bastian 2006). By scaling up this result, massive galaxies are
expected to reach a new equilibrium  after $\sim 2\,$Gyr, at $t_{\rm int}\sim
t_{\rm QSO}+2\,$Gyr (asterisks in Fig.~\ref{fig:re}). Since then only minor
adiabatic mass losses occur, producing slight changes of radius and velocity
dispersion (from asterisks to squares). We recall that the quasar activity
statistically reaches its maximum at redshift $z\sim 2$, corresponding to a
cosmic time $t_{\rm cosm}\sim 3\,$Gyr. The rapid mass loss is then expected to
statistically peak at the same redshift, triggering size variations that
stabilize after $\sim 2\,$Gyr. Thus the massive galaxies should on the average
reach the local size-mass relation at $t_{\rm cosm}\sim 5\,$Gyr, corresponding
to $z\sim 0.8$. This result is in keeping with sizes observed at $z \lesssim 1$
(Cimatti et al. 2008) and with results of studies of the Fundamental Plane at
$z\leq 0.8$ (Renzini 2006). In the redshift interval $0.8 \leq z \leq 3$ a
large scatter in the size--stellar mass relation is expected and indeed
observed.

This scenario may appear to be contradicted by observations of
Ferguson et al. (2004), who found an average size $r_e\simeq
1.7\,$kpc for luminous Lyman Break Galaxies (LBGs) at $z\simeq
4$, while we  would expect  $r_e\sim 0.5\,$kpc, inserting the
appropriate masses in eq.~(\ref{eq:re}). However as pointed out
by Joung, Chen \& Bryan (2008), the apparent effective radius
could be a factor of about 3 larger than the intrinsic one,
because of the presence of dust in the central star forming
regions.

In the case of rapid mass loss, the velocity dispersion of the
stars at virial equilibrium scales as $\sigma_{\star}\propto
r_e^{-1/2}$. As shown by Fig.~\ref{fig:sigma}, the size
evolution illustrated by Fig.~\ref{fig:re} brings the quite
high velocities inferred from the observed values of $M_\star$
and $r_e$ in the compact phase, to within the locally observed
range.

The structural evolution of low mass E galaxies follows a different path,
driven by the slow mass loss due to galactic winds and supernova explosions,
because their nuclear activity is of low power. Adopting a Chabrier (2005)
Initial Mass Function (IMF), we find a mass loss of a factor of $\sim 2$ on a
time scale of several Gyrs (the factor is only $\sim 1.4$ for a Salpeter (1955)
IMF); after eq.~(\ref{eq:slow}) the size increases by the same factor. However
most of the expansion occurs when these galaxies are young. For a Chabrier IMF,
already half a Gyr after the baryon collapse the size has increased by a factor
of $\sim 1.6$ (triangles in Fig.~\ref{fig:re}) and the subsequent expansion is
limited to a factor of $\sim 1.3$ (squares). Thus such galaxies observed at
ages $\gtrsim 0.5\,$Gyr should exhibit a size quite close to that of local
spheroidal galaxies with the same stellar mass. As a consequence, high-$z$
galaxies with $M_\star < 2\times 10^{10}\,M_{\odot}$ should also exhibit a
smaller scatter in the effective radius--stellar mass relation. They are
predicted to have significantly more compact sizes only at ages $\lesssim
0.5\,$Gyr, when their star formation rates are of tens $M_\odot$/yr (see Fig.~1
of Lapi et al. 2006), typical of relatively low luminosity, high-$z$ LBGs. In
the adiabatic expansion case the velocity dispersion scales as
$\sigma_{\star}\propto r^{-1}$. Again, this brings the high initial velocities
within the locally observed range.

As apparent from Fig.~\ref{fig:re}, we predict significant evolution for
galaxies with $M_{\star}\geq 2\times 10^{10}$ M$_{\odot}$, well below the
threshold of $M_{\star}\geq 5\times 10^{11}\,M_{\odot}$ implied by the
semi-analytic model of Khochfar \& Silk (2006).

In conclusion we suggest that the rapid mass loss driven by the
quasar feedback is the main agent of the size and velocity
dispersion evolution of massive spheroidal galaxies. Lower-mass
galaxies experience a weaker, but non negligible evolution (of
amplitude depending on the adopted IMF), due to mass loss
mainly powered by supernova explosions; most of it occurs
during their active star-formation phase. Although our
calculations have been carried out in the framework of the
Granato et al. (2004) model, this evolutionary behaviour is a
generic property of all models featuring large mass loss due to
quasar and/or supernova feedback. Observational evidences, some
of which are briefly summarized in \S\,\ref{sect:size}, suggest
that quasar driven high velocity outflows may have removed from
the central regions of massive galaxies a gas mass comparable
to the mass in stars on a timescale shorter than the dynamical
time. If so, simple, model independent, physical arguments,
presented in \S\,\ref{sect:loss}, imply a swelling of the
stellar distribution and a decrease of the stellar velocity
dispersion. The model contributes detailed, testable,
predictions on the evolution of the effective radius and of the
velocity dispersion as a function of the galactic age and of
the halo mass. It specifically predicts different evolutionary
histories for galaxies with present day stellar masses above
and below $M_{\star}= 2\times 10^{10}\, M_{\odot}$.

The energy injected by dry mergers into the stellar systems can also enlarge
their size. However mergers increase the mass in stars too, and, to first
order, move galaxies roughly parallel to the size-mass relation (van Dokkum et
al. 2008; Damjanov et al. 2008), while the data show size evolution at fixed
mass in stars.

The simple analysis presented in this {\it Letter} is intended
to be a first exploration, sketching a promising scenario for
interpreting challenging observational results. This scenario
can be tested, on one side, by numerical simulations with
appropriate time resolution, properly taking into account the
interactions between baryons and DM, and, on the other side, by
observations of size and velocity dispersion especially of
lower mass high-$z$ galaxies, that are expected to show an
evolutionary behaviour different from that of massive galaxies
because of the different mass loss history.

\begin{acknowledgements}
This research has been partially supported by ASI contract
I/016/07/0 ``COFIS''.
\end{acknowledgements}

\end{document}